\documentclass[onecolumn,showpacs,preprintnumbers,amsmath,amssymb,10pt]{revtex4}

\usepackage{epsf}
\usepackage{graphicx}  % Include figure files
\usepackage{dcolumn}   % Align table columns on decimal point
\usepackage{bm}        % bold math

\newcommand{\be}{\begin{equation}}
\newcommand{\en}{\end{equation}}
\newcommand{\bea}{\begin{eqnarray}}
\newcommand{\ena}{\end{eqnarray}}
\newcommand{\cp}{Chaplygin gas}

\newcommand{\nn}{\nonumber}
\newcommand{\g}{\gamma}

\begin{document}

\title{Hybrid Chaplygin gas}

\author{Hongsheng Zhang}

 \affiliation{Department of Astronomy, Beijing Normal University,
Beijing 100875, China}
\affiliation{
 Korea Astronomy and Space Science Institute,
  Daejeon 305-348, Korea }
 \author{Zong-Hong Zhu\footnote{E-mail address: zhuzh@bnu.edu.cn}}
\affiliation{
 Department of Astronomy, Beijing Normal University, Beijing 100875, China}

 \author{Lihua Yang}
\affiliation{School of Network Education, Beijing University of
Posts and Telecommunication, Beijing 100088, China}

\date{ \today}

\begin{abstract}

 Hybrid Chaplygin gas model is put forward, in which the gases ~play
 the role of dark energy. For this model the coincidence problem is greatly alleviated.
 The effective equation of state of the dark energy may cross
 the phantom divide $w=-1$.
 Furthermore, the crossing behaviour is
  decoupled from any gravity theories. In the present model, $w<-1$
  is only a transient behaviour. There is a
  de Sitter attractor in the future infinity. Hence, the big rip singularity, which often
  afflicts the models with matter whose effective equation of state less than $-1$,
  is naturally disappear. There exist stable scaling solutions, both
  at the early universe and the late universe. We discuss the
  perturbation growth of this model. We find that the index is consistent
  with observations.

\end{abstract}

\pacs{ 95.36.+x  } \keywords{\cp, dark energy}

\maketitle

\section{Introduction}

 The existence of dark energy is one of the most significant cosmological discoveries
 over the last century \cite{acce1,acce2}. Although fundamental for
our understanding of the Universe, its nature remains a completely
open question nowadays. For recent studies of dark energy, see
 review article \cite{review}.

 Recently the so-called \cp, also dubbed quartessence, was suggested as a
candidate of a unified model of dark energy
 and dark matter \cite{cp}.  The \cp~ is characterized by an exotic equation
of state (EOS)
 \be
 p_{ch}=-\tilde{A}/\rho_{ch},
 \label{state}
 \en
 where $\tilde{A}$ is a positive constant. The above equation of state leads to
 a density evolution in the form
 \be
 \frac{ \rho_{ch}}{\rho_0}=\sqrt{A+B(1+z)^6},
 \label{evo}
 \en
 where $A\equiv \tilde{A}/\rho_0^2$, $B$ is an integration constant, $z$ denotes the redshift,
 $\rho_0$ represents the present critical density.
The attractive feature of the model is that it naturally unifies
 dark energy and dark matter. The reason is that, from
(\ref{evo}), the \cp~ behaves as dust-like matter at early stage and
as a cosmological constant at later stage. Some possible motivations
for this model from the field theory points of view are
 investigated in ~\cite{fields1,fields2}. The Chaplygin gas emerges as an effective
fluid associated with $d$-branes~\cite{brane,brane1} and can be also
 obtained from the Born-Infeld action~\cite{bi}.

  The \cp~ model has been thoroughly investigated
 for its impact on the 0th order
cosmology, i.e., the cosmic expansion history (quantified by the
Hubble parameter $H[z])$ and corresponding spacetime-geometric
observables. An interesting range of models was found to be
consistent with SN Ia data \cite{sn1,sn2,sn3,sn4}, CMB peak
locations \cite{cmb} and gas mass fractions in clusters of
galaxies\cite{zhzh}. There seems to be, however, a flaw in unified
dark matter (UDM) models that manifests
 itself only on small (galactic) scales and that has not been revealed by the
 studies involving only background tests. In \cite{antiudm},
  it is found that generalized \cp~  (GCG) model produces oscillations or exponential blowup of the matter
   power spectrum inconsistent with observations. In fact, from this analysis, 99.999 \% of
   previously allowed parameter of GCG model has been excluded (see, however, \cite{bbc1,bbc2,bbc3}).

            Hence we may turn to a model with \cp~and dark matter.
  It has been pointed out that \cp~ model can be described by
 a quintessence filed with well-connected potential \cite{cp}.
 Therefore, a model with \cp~ and dark matter
 is essentially a special quintessence model, in which the EOS of dark energy
 (\cp) $w$ (defined as the ratio of
 pressure to energy density)
 satisfies $0<w<-1$.

  A merit of the \cp~ model, in which \cp ~only plays the role
 of dark energy, is that the coincidence problem is greatly
 alleviated. The coincidence problem in $\Lambda$CDM model says that
 $\Lambda$ keeps a constant while the density of CDM evolves as $(1+z)^3$
 in the history of the universe, then why do they approximately equal
 each other at ``our era"? We see from (\ref{evo}) that density of
 \cp~ evolves as the same of dark matter and only at late time it
 evolves as cosmological constant. Therefore, the coincidence
 problem may be alleviated at some degree. We shall discuss this
 possibility.

   The cosmological constant is the far simple candidate for dark energy.
   Though previous observations are consistent with the cosmological constant
    , they leave enough space for a dynamical dark energy \cite{antiv1,antiv2,antiv3,WMAP}.
    Following the more accurate data, the implications for dynamical dark
    energy become clear:
  the recent analysis of the type Ia supernovae data
  indicate that the time varying dark energy gives a better
  fit  than a cosmological constant,
  and in particular, the equation of state parameter
   $w$  crosses $-1$ at some low redshift region from above to below
 \cite{vari1,vari2,vari3,vari4,vari5, cross1,cross2,cross3,cross4,cross5,cross6,cross7,
 cross8,cross9,cross10,cross11,cross12,cross13,cross14,cross15,cross16,cross17,cross18,cross19,
 cross20,cross21,cross22,cross23,cross24,cross241,cross25,cross26,cross27}. It seems difficult to realize this
 transition in context of \cp~ dark energy model. However, if we
 consider EOS (1) carefully and rigorously, we shall find the other
 solution of the continuity equation different from (2),
 \be
 \label{minus}
   \frac{\tilde{\rho}_{Ch}}{\rho_0}=-\sqrt{C+D(1+z)^6},
 \en
 which satisfies the EOS,
 \be
 p_{Ch}=-\frac{C\rho_0^2}{\rho_{Ch}},
 \en
 where $C$ and $D$ are constants. A noticeable property of this case is that the
 energy density of \cp~ is negative. But this is not anything
 completely new. If quantum effect is considered, the energy density
 of a field can be naturally lower than zero \cite{bd}. Also in the
 phantom dark energy model which is extensively studied in context of
 cosmology \cite{ph}, the density of dark energy can be less than zero if observed by
 observers other than the homogeneous observers, since it violates
 the weak energy condition. We call this \cp~ with negative energy
 density ``type II" \cp~, and correspondingly, the original \cp~ with
 positive density is dubbed ``type I" \cp~ in the present article.
 Therefore, we find there are two branches of mathematical solution satisfying the EOS
 of \cp~and continuity equation. In fact, $B$ in (2), as an
 integration constant, can also be negative. However, a difficult
 arises when $B<0$: the universe bounces at some finite redshift. To
 avoid to plague the success of nucleosynthesis, $|B/A|<6\times
 10^{-28}$, which is an unnatural fine-tuning. In this paper we
 constrain ourself in the case of $B>0$.

  A kind of interacting \cp ~model in which the \cp
 ~plays the role of dark energy and interacts with cold dark matter
  particles has been investigated in \cite{self}.
  In this model the effective equation of state of \cp~ may cross the
   phantom divide.

   We shall present a hybrid model composed by both of the two
   types of \cp, in which the EOS of dark energy can cross the
   phantom divide, which is decoupled from any gravity theory.
   Assuming standard general relativity, we prove there exist
   scaling solutions, both in the early universe and the late time
   universe: the early universe is attracted by a dust tractor, that
   is, the cosmic fluids enter a dust-like phase; and  the late
   time universe is attracted by a de Sitter tractor
   , that is, it evolves into a de Sitter phase.

   Often, the big rip singularity is narrated as follow: at finite
   cosmic time, both the scale factor and energy density become
   infinite\cite{rip}. This is a conclusion in a special chat, although in the
   most used chat. As we learned from Schwarzschild solution, an
   ordinary point can be a singularity in disguise for a special
   chat. We shall present some  discussions on this issue and
   show the big rip singularity is a true singularity. The big rip
   singularity usually emerges at the models with phantom like dark
   energy, because the density of a phantom field will increase with
   time. But in our present model $w<-1$ is only a transient
   behavior, and the final state of the universe is always a de
   Sitter, for the whole parameter space.

  To
 construct a model simulating the accelerated expansion is not very
 difficult. That is the reason why we have so many different models.
   Recently, some suggestions are presented that perturbation
   (fluctuation) growth
 function $\delta(z)\equiv\delta\rho_m/\rho_m$ of the linear matter
 density contrast as a function of redshift $z$ can be an effective probe to
  explore the dark energy models \cite{growth1,growth2,growth3,growth4,growth5,growth6,growth7,growth8,growth9,growth10,growth11,growth12,growth13,growth14}. The related
  works can be traced back to \cite{ratra1,ratra2,ratra21,ratra3}.

  There is an  approximate relation between the growth function
  and the partition of dust matter \cite{growth1,growth2,growth3,growth4,growth5,growth6,growth7,growth8,growth9,growth10,growth11,growth12,growth13,growth14},
   \be
   \label{f}
 f\equiv\frac{d\ln\delta}{d\ln a}=\Omega_m^\gamma,
     \en
 where $\Omega_m$ is the density partition of dust matter, $a$
 denotes the scale factor, and $\g$ is the growth index. This relation is a
 perfect approximation at high redshift region. Also, it  can be
 used
 in low redshift region, see for example \cite{growth1,growth2,growth3,growth4,growth5,growth6,growth7,growth8,growth9,growth10,growth11,growth12,growth13,growth14}.  The
 theoretical value of $\g$ for $\Lambda$CDM model is $6/11$
 \cite{ratra1,ratra2,ratra21,ratra3}. We shall investigate the perturbation growth of the
 present model.

     We present our model in details in the next section. In section
     III, we study the perturbation growth in the hybrid \cp~ model.
     Our conclusions and discussions appear in the last section.

\section{the hybrid model }
 In this section we shall investigate three properties of the hybrid
 \cp~ model. In the first subsection we display that the coincidence
 problem is greatly alleviated. In subsection B, we study the crossing
  $-1$ behavior of the EOS of the dark energy. And we explore the
  the evolution of this model by the dynamical-system analyses in subsection C.
\subsection{Coincidence problem}
 We consider a model in which the type I \cp~ and type II \cp~ play
 the role of dark energy together. The density of dark energy
 reads
 \be
 \rho_{de}=\rho_1+\rho_2.
 \en
 The continuity equation for which reads,
 \be
 {d\rho_{de}}=3(\rho_{de}+p_{de})d\ln(1+z).
 \label{conti}
  \en
  Here $\rho_1$ denotes the density of type I \cp~ and $\rho_2$
  represents the density of type II \cp ~,
  \bea
  \label{rho1}
  \frac{\rho_1}{\rho_0}&=&\sqrt{A+B(1+z)^6},\\
  \frac{\rho_2}{\rho_0}&=&-\sqrt{C+D(1+z)^6}.
  \label{rho2}
  \ena
  The energy density of dust matter redshifts as
  \be
  \frac{\rho_m}{\rho_0}=\Omega_{m0} (1+z)^3.
  \en
 Hence, if dark energy is just cosmological constant, we suffer from
 a coincidence problem. However, if \cp~plays the role of dark
 energy, this problem will be alleviated because at the early
 universe the \cp~ also behaves as dust.

 \begin{figure}
 \centering
 \includegraphics[totalheight=2.6in, angle=0]{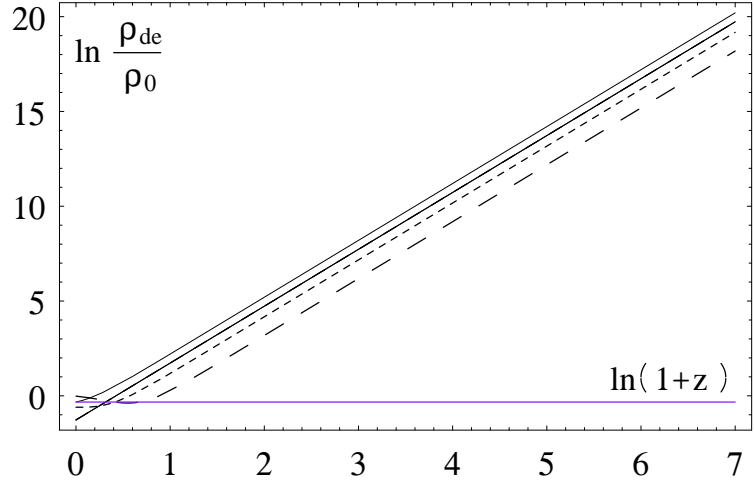}
 \caption{A double logarithmic plot: the densities of dark energy as
functions of redshift. The cosmological constant inhabits on the
horizontal line, the dust matter dwells on the thick solid curve,
 and other three curves represent the \cp~ dark energy. The long dashed, short dashed, and thin solid
 curves
 denote the dark energy in a positive curvature, negative curvature and flat universe, respectively.}
 \label{coin}
 \end{figure}
 We see from figure \ref{coin} that if a cosmological constant
 plays the role of dark energy, the density of matter is about 20
 orders larger than the dark energy at $z=1100$: they must have been fine-tuned
  at that time. However, if the \cp~ serves as dark energy, the ratio
 of dark matter and dark energy keeps at order 1, which relieves the
  coincidence problem, for all 3 cases of the universe.

 \subsection{Crossing $-1$}
\begin{figure}
 \centering
 \includegraphics[totalheight=2.6in, angle=0]{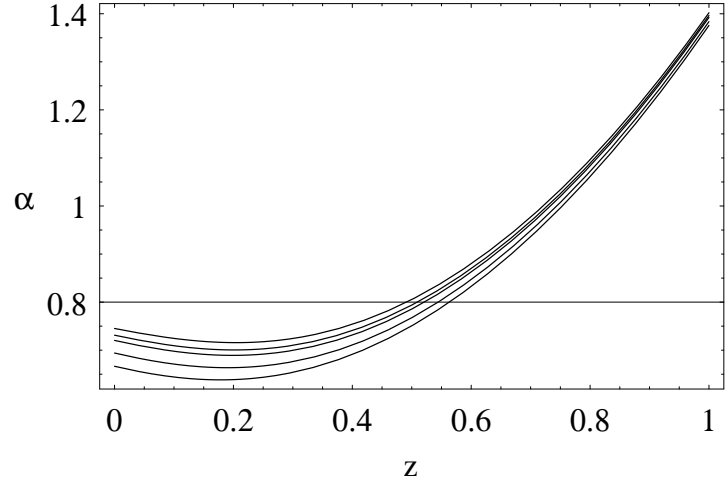}
 \caption{$\alpha$ as a function of $z$. On all of the 5
 orbits $B=0.3,~D=0.15,~ \Omega_m=0.28$ with different $A$ and $C$.
 From the above to the below $A=1.104,C=0.048;~~A=1.059,C=0.039;~~A=1.01,~C=0.03;~~
 A=0.9106,C=0.015;~~A=0.8267,C=0.006$}
 \label{rhoz}
 \end{figure}

\begin{figure}
\centering
\includegraphics[totalheight=2.6in, angle=0]{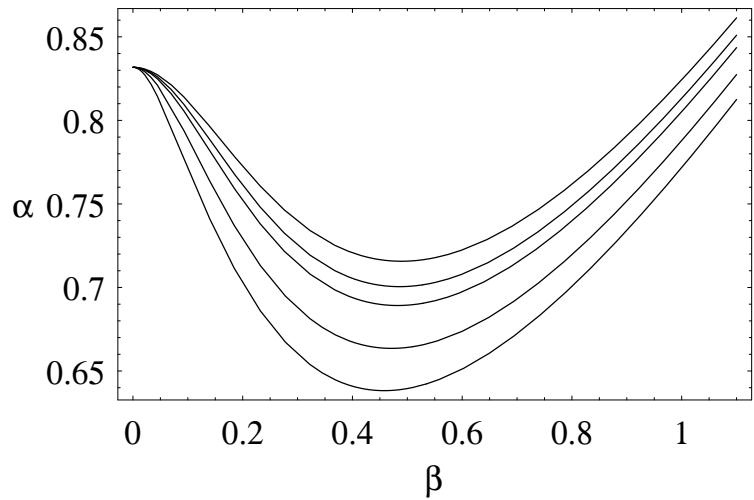}
\caption{Phase portrait for $\alpha$ vs. $\beta$. On all of the 5
 orbits $B=0.3,~D=0.15,~ \Omega_m=0.28$ with different $A$ and $C$.
 From the above to the below $A=1.104,C=0.048;~~A=1.059,C=0.039;~~A=1.01,~C=0.03;~~
 A=0.9106,C=0.015;~~A=0.8267,C=0.006$}
 \label{portrait}
 \end{figure}

  From the continuity equation (\ref{conti}), we arrive at
      \be
  w_{de}=\frac{p_{de}}{\rho_{de}}=-1-\frac{1}{3}\frac{d \ln \rho_{de}}{d \ln
  (1+z)},
   \en
   which means that in an expanding universe
   if $\rho_{de}$ decreases and then increases
  with respect to the redshift, or increases and then
  decreases, then we conclude that EOS of dark energy crosses phantom
  divide. Here,
  \be
  \frac{d\rho_{de}}{dz}=-3[B\rho_1^{-1} (1+z)^5+D\rho_2^{-1}
  (1+z)^5].
  \en
  We see that if we carefully tune $B$ and $D$, the EOS of dark
  energy may cross $-1$. In figure \ref{rhoz}, we show some concrete
  examples for which \cp~ crosses the phantom divide at about $z=0.2$, where
  $\alpha\triangleq\rho_{de}/\rho_0$.
   Also, in figure \ref{portrait} we show the phase
  portrait of $\alpha$ vs. $\beta\triangleq\rho_m/\rho_0$,
  where $\rho_m$ represents the density of dust matter. A note is that
 any gravity theory is not interposed up to now. Our results only
 depend on the continuity equation.

 \subsection{Dynamical analysis}

  Often, a model containing phantom-like matter is afflicted by big
  rip problem, that is, at some finite time, the scale factor and density
  is divergent. However, this description of singularity depends on
  FRW coordinates, which may be an ordinary point described by other coordinates.
  We check scalar polynomial curvature, which is a good lesson we learned from Schwarzschild solution. Here we take the example in
  \cite{guophan}. Consider a universe is dominated by dust and then by constant $w$ phantom. The
  transition occurs at $t_{pm}$. At
  \be
  t=\frac{wt_{pm}}{1+w},
  \en
 the scale factor and density become divergent. The simplest
  scalar polynomial is Ricci scalar $R$,
  \be
  R=\left[\frac{16}{3t_{pm}^2}-\frac{4}{t_m^2}(1+w)\right]\left[-w+(1+w)\frac{t}{t_{pm}}\right]^{-2}.
  \en
  Because $w<-1$, the term in first square bracket always larger than zero, and term in the
  second bracket is divergent when $t\to \frac{wt_{pm}}{1+w}$.
  Therefore, the scalar curvature is infinite when it goes to the
  big rip point. Also, because all comoving observers move along timelike geodesics, and the cosmic time is
  their proper time, this singularity is also an incomplete-geodesic singularity. Hence, big rip singularity
  is a true singularity. Through researches on the dynamical properties
  of the universe with \cp~ and dark matter, we shall show that there is
  no future singularity in this model, though the dark energy
  behaves  as phantom in some stage.

 Now we start to study the dynamical evolution of the universe in frame of standard
 general relativity, for which we introduce Friedmann equation in a spatially flat
 FRW universe, which is implied
 either by theoretical side (inflation in the early universe)
 ,or observation side (CMB fluctuations \cite{WMAP}),
 \be
 H^2=\frac{1}{3\mu^2}(\rho_1+\rho_2+\rho_m)
 \label{fried}
 \en
 where, as usual, $H$ denotes the Hubble parameter, and $\mu$ stands
 for the reduced Planck mass. For convenience we first define the following new
  dimensionless variables,
  \bea
  x&\triangleq&\frac{\sqrt{\rho_1}}{\sqrt{3}\mu H},\\
  y&\triangleq&\frac{\sqrt{\rho_2}}{\sqrt{3}\mu H},\\
  u&\triangleq&\frac{\sqrt{\rho_m}}{\sqrt{3}\mu H},\\
  l_1&\triangleq&\frac{A^{1/4}}{\sqrt{3}\mu H},\\
  l_2&\triangleq&\frac{C^{1/4}}{\sqrt{3}\mu H}.
  \ena
  The dynamics of the universe can be described by the following
  dynamical system with these new dimensionless variables,
 \bea
 \label{1}
 x'&=&-\frac{3}{2}
 x^{-1}(x^2-x^{-2}l_1^{4})+\frac{3}{2}xP,\\
 \label{2}
  y'&=&-\frac{3}{2}
 y^{-1}(y^2-y^{-2}l_2^{4})+\frac{3}{2}yP,\\
 \label{3}
   u'&=&-\frac{3}{2}u+\frac{3}{2}uP,\\
 \label{4}
  l'_1&=&\frac{3}{2}l_1P,\\
  \label{5}
 l'_2&=&\frac{3}{2}l_2P,
 \ena
 where
 \be
 P=x^2-x^{-2}l_1^4-(y^2-y^{-2}l_2^4)+u^2,
 \label{P}
 \en
  and a prime stands for derivation with respect to
 $s\triangleq-\ln(1+z)$. Note that the 4 equations (\ref{1}), (\ref{2}),
 (\ref{3}), (\ref{4}), (\ref{5}) of this system are not independent. By using the Friedmann
 constraint, which can be derived from the Friedmann equation,
 \be
 \label{constraint}
 x^2-y^2+u^2=1,
 \en
 the number of the independent equations can be reduced to 4.
    The
  critical points of this system satisfying $x'=y'=l'=b'=0$ appearing at
 \bea
 \label{11}
 l_1=l_2=0,\\
 \label{12}
 x^2-y^2+u^2=1,
 \ena
 and
 \bea
 \label{21}
 u=0,\\
 \label{22}
 x=l_1,\\
 \label{23}
 y=l_2.
 \ena
  The first set of critical points (\ref{11}), (\ref{12}) dwells at the early
  universe, since $H\to \infty$. Also this set satisfies the
  constraint equation automatically. One sees that it is fairly
  ample, which only needs $(x_c,y_c,u_c)$ inhabits on the surface
  (\ref{12} ), where $c$ label the critical point. The reason roots in the fact that
  all of the three components, type I \cp, type II \cp~ and matter
  are dust-like in the early universe, they can evolve into a scaling
  solution with rather arbitrary proportion of components along inverse time direction.
   We call
  this set of critical points dust attractor.
        The second set of critical points resides at the late
  time universe, because $u=0$, which means matter has been
  infinitely diluted. By using constraint equation
  (\ref{constraint}), we further derive $x_c-y_c=1$, which resides on a
  hyperbola. Because this set of critical points ensure that
  $l_1=l_2=$~constant, which means $H=$~constant, we call it de Sitter
  attractor.
   The previous models in which the EOS of dark energy
  crosses $-1$ often suffer from future or past difficulties, such
  as big rip disaster or to plague the structure formation theory
  because of a too stiff EOS. Hence, most of them only can be used
  to describe the evolution of the universe at some low redshift. By
  striking contrast, our model are free of such difficulties from CMB
  decoupling to the future infinity. Global fittings or simulations
  of structure formation operate routinely in frame of the present
  model.

  Though in above context we call the singularity ``attractor", it
  is only an intuitive conclusion. To obtain a mathematically strict
  result of the stability at the neighborhood of the singularities,
  imposing a perturbation to the system up to the linear order, we
  obtain

  \bea
  \nn
  (\delta x)'&=&
  \left(-\frac{3}{2}-\frac{9l_1^4}{2x^4}+\frac{3}{2}P+3x^2+
  \frac{3l_1^4}{x^2}\right)\delta x\\
  \nn
       &-&\left(3xy+3\frac{l_2^4x}{y^3}\right)\delta
       y+3xu\delta u\\
       &+&\left(\frac{6l_1^3}{x^3}-\frac{6l_1^3}{x}\right)\delta l_1
       +\frac{6l_2^3x}{y^2}\delta l_2,\\
   \nn
  (\delta y)'&=&
  \left(-\frac{3}{2}-\frac{9l_2^4}{2y^4}+\frac{3}{2}P-3y^2-
  \frac{3l_2^4}{y^2}\right)\delta y\\
  \nn
       &+&\left(3xy+3\frac{l_1^4y}{x^3}\right)\delta
       x+3yu\delta u\\
       &+&\left(\frac{6l_2^3}{y^3}+\frac{6l_2^3}{y}\right)\delta l_2
       +\frac{-6l_2^3y}{x^2}\delta l_1,\\
  \nn
  (\delta u)'&=& \left(-\frac{3}{2}+\frac{3}{2}P+3u^2\right)\delta
  u \\
  \nn
  &+& \left(3ux+3\frac{ul_1^4}{x^3}\right)\delta
  x+\left(-3uy-3\frac{ul_2^4}{y^3}\right)\\
  &-&6\frac{ul_1^3}{x^2}\delta l_1+6\frac{ul_2^3}{y^2}\delta l_2,\\
  \nn
  (\delta l_1)'&=&
  \left(\frac{3}{2}P-6\frac{l_1^4}{x^2}\right)\delta
  l_1+\left(3xl_1+3\frac{l_1^5}{x^3}\right)\delta x\\
  \nn
    &+& \left(-3l_1y-3\frac{l_2^4l_1}{y^3}\right)\delta y
    +6\frac{l_2^3l_1}{y^2}\delta
  l_2\\
  &+&3l_1u\delta u,\\
  \nn
  (\delta l_2)'&=&
  \left(\frac{3}{2}P+6\frac{l_2^4}{y^2}\right)\delta
  l_2+\left(-3yl_2-3\frac{l_2^5}{y^3}\right)\delta y\\
  \nn
    &+& \left(3l_2x+3\frac{l_1^4l_2}{x^3}\right)\delta x
    -6\frac{l_1^3l_2}{x^2}\delta
  l_2\\
  &+&3l_2u\delta u,
  \ena
 where $P$ is defined in (\ref{P}). Before calculation of the eigenvalues
 of the linearized system around singularities, we must point out a key difference
 between the two singularities. The criterion for the stability of a small deviation
 must be with an argument which increases when the orbits go to the
 singularity. It is easy to check that the argument $s=-\ln (1+z)$ in (\ref{1})-(\ref{5})
  is adapted
 to the attractor in late time universe.  The dust attractor resides at the early universe,
 while the de Sitter attractor inhabits in late time universe. Thus,
 the linearized system about them should be described by arguments
 with opposite sign. If we use the same argument $s$, a positive
 definite eigenmatrix implies that the dust attractor is stable.

    For the dust attractor in the
 early universe, substitute the variables by the values given in
 (\ref{11}), (\ref{12}), we obtain the eigenvalues of the linearized
 system
 \be
 \nn
 \lambda_1=0,~\lambda_2=0,~\lambda_3=3,~\lambda_4=\frac{3}{2},~
 \lambda_5=\frac{3}{2}.
 \en
 Hence, base on the above discussions, the dust attractor is
 quasi-stable.
 For the de Sitter attractor in late time universe,
 substitute the variables by the values given in
 (\ref{21}), (\ref{22}), (\ref{23}), we obtain the eigenvalues of the linearized
 system
 \be
 \nn
 \lambda_1=-6,~\lambda_2=-6,~\lambda_3=0,~\lambda_4=0,~
 \lambda_5=-\frac{3}{2}.
 \en
 which is also a
 quasi-stable attractor.

 Hence, eventually, the universe enters a de Sitter phase, which
 does not contain any singularities, which means the big rip disaster disappears naturally.
 The phantom behavior of the
 dark energy is only a transient phenomena. Also this result does
 not depend on parameter selection and initial condition because of attractor
 behavior.

\begin{figure}
\centering
\includegraphics[totalheight=2.6in, angle=0]{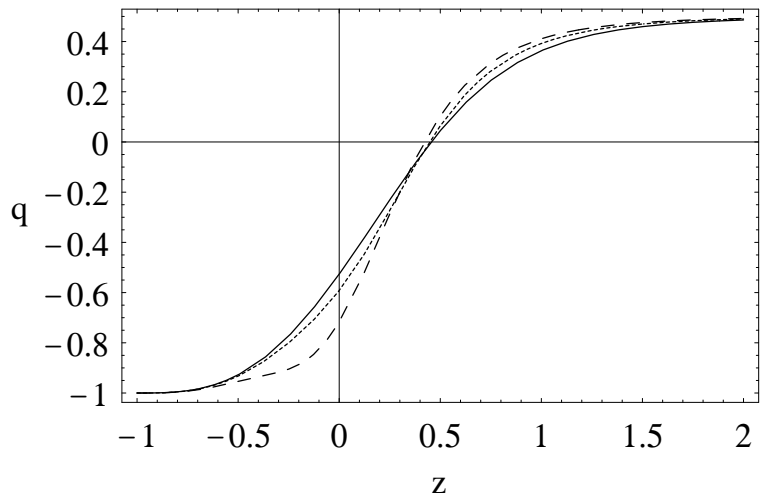}
\caption{Deceleration parameter $q$ as a function of $z$. On all of
the curves orbits $B=0.3,~D=0.15,~ \Omega_m=0.28$ with different $A$
and $C$. For the solid curve $A=5.92,~C=3.00$, the short dashed
curved $A=1.63,~C=0.300$, while for the long dashed curve
$A=1.01,~C=0.0300$. }
 \label{dece}
 \end{figure}

  The most significant parameter from the viewpoint of
  observations is the deceleration parameter $q$, which carries the total
  effects of cosmic fluids. Here $q$ reads
  \be
  q=-\frac{\ddot{a}a}{\dot{a}^2}=\frac{1}{2}\left[1-3(\frac{A}{\rho_1}+\frac{C}{\rho_2})
  \frac{1}{\rho_1+\rho_2+\rho_m}\right],
  \en
  and density of \cp~ $u$ and density of dark matter $v$ should
  satisfy
  \be
  x^2(0)-y^2(0)+u^2(0)=1.
  \label{con}
  \en
  And then Friedmann equation ensures the spatial flatness in the
  whole history of the universe. Note that not all the examples in
  figure 1, 2 or 3 satisfies this constraint, since we do not introduce
  Friedmann equation there.
  Before analyzing the evolution of $q$ with redshift,
   to obtain some asymptotic behaviors of the universe is useful. When $z\to \infty$, $q$
 must go to $1/2$ because both type I, II \cp~ and matter behave as dust
 ; while when $z\to -1$, $q$ is determined by
 \be
  \lim_{z\to -1} q= \lim_{z\to
  -1}\frac{1}{2}[1-3(x^{-2}l_1^4-y^{-2}l_2^4)]=-1,
  \en
  which agrees with the above analysis of the dynamical properties of this system.
  Here we carefully
  choose a new set of parameter which satisfies Friedmann constraint
  (\ref{con}), and plot the deceleration parameter $q$ in figure
  \ref{dece}. We see for these three set of parameters the
  deceleration parameters are well consistent with observations.

%%%%%%%%%%%%%%%%%%%%%%%%%%%%%%%%%%%%%%%%%%%%%%%%%%%%%%%%%%%%%%%
\section{perturbation growth}

 The Friedmann equation in the present model is given by
 (\ref{fried}). Now we consider the perturbation growth in this hybrid model.
  After the matter decoupling from radiation, for a region well inside a Hubble radius, the perturbation growth
  satisfies the following equation \cite{liddle},
  \be
  \ddot{\delta}+2H\dot{\delta}-\frac{1}{6\mu^2}\rho_m \delta=0.
  \label{pertu}
  \en

  With the partition functions,
  \bea
  \Omega_m &=& \frac{\rho_m}{3\mu^2 H^2},\\
   \Omega_{de}&=&\frac{\rho_{de}}{3\mu^2 H^2},
      \ena
 the perturbation equation (\ref{pertu}) becomes,
  \be
   \left(\ln\delta\right)''
 +\left(\ln\delta\right)'^{2}
 +\left(2+\frac{H'}{H}\right)\left(\ln\delta\right)'
 =\frac{3}{2} \Omega_m,
  \label{f1}
  \en
  where a prime denotes the derivative with respect to $\ln a$.
  $\Omega_m$ and $\Omega_{de}$ evolve as
  \bea
  \Omega_m  =\frac{ \Omega_{m0}(1+z)^3}{\Omega_{m0}(1+z)^3+\sqrt{A+B(1+z)^6}-\sqrt{C+D(1+z)^6}}, \\
  \Omega_{de} = \frac{\sqrt{A+B(1+z)^6}-\sqrt{C+D(1+z)^6}}{\Omega_{m0}(1+z)^3+\sqrt{A+B(1+z)^6}-\sqrt{C+D(1+z)^6}},
  \ena
  where 0 denotes the present value of a quantity.
  The growth function defined in (\ref{f}) is just $({\ln a})'$.
  Thus (\ref{f1}) generates,
  \be
  f'+f^2+\left[\frac{1}{2}(1+\Omega_k)+\frac{3}{2}w_{de}(\Omega_m+\Omega_k-1)\right]f=\frac{3}{2}\Omega_m,
  \label{diff}
  \en
  where we have used

    \be
  \frac{H'}{H}=-\Omega_k-\frac{3}{2}\left[\Omega_m+(1+w)\Omega_{de}\right],
  \en
  and
 \be
  \Omega_m+\Omega_{de}=1.
 \en

 Recalling (\ref{f}), we derive the evolution equation for $\g$
 from (\ref{diff}),
    \be
    3w_{de}(1-\Omega_m)\ln \Omega_m \frac{d\g}{d\ln
    \Omega_m}+\Omega_m^{\g} -\frac{3}{2}\Omega^{1-\g}
    +3w_{de}\Omega_m(1/2-\g)+3\g w_{de}-\frac{3}{2} w_{de} +1/2=0,
    \en
  where we have used
  \be
  (\ln \Omega_m)'=3(1-\Omega_m)w_{de}.
  \en
  By using (\ref{rho1}) and (\ref{rho2}), we obtain
  \be
  w_{de}=-\left[\frac{A}{\sqrt{A+B(1+z)^6}}-
  \frac{C}{\sqrt{C+D(1+z)^6}} \right] \left[ {\sqrt{A+B(1+z)^6}}-
  \sqrt{C+D(1+z)^6} \right]^{-1}.
  \en
  Expanding $\g$ about $1-\Omega_m$, we get
  \be
  \g=\frac{3}{5-w_{de}/(1-w_{de})}+2.4\times
  10^{-2}(1-\Omega_m)(1-w_{de})(1-3w_{de}/2)(1-6w_{de}/5)^{-3}.
  \en
  Here we present some examples to illuminate the current numerical value of
  $\g$ in the hybrid \cp~ model. We take the same examples as in
  figure 4. $B=0.3,~D=0.15,~ \Omega_m=0.28$ with different $A$
  and $C$. For $A=5.92,~C=3.00$, $\g=0.555$; for $A=1.63,~C=0.300$,
  $\g=0.553$;  for $A=1.01,~C=0.0300$, $\g=0.550$. The theoretical
  values of this model are well consistent with observations
  \cite{growth1,growth2,growth3,growth4,growth5,growth6,growth7,growth8,growth9,growth10,growth11,growth12,growth13,growth14}.

 \section{Conclusion and discussion}
  Through careful analysis, we find a new kind of \cp~, called type II \cp, whose
  energy density is negative. Then we present a hybrid \cp~ model,
  in which type I \cp~and type II \cp~play the role of dark energy
  together. The EOS of dark energy crosses $-1$ naturally without
  introducing any gravity theories.

  In frame of standard general relativity and a spatially flat FRW universe,
  we study the dynamical properties of the present model. We find
  attractor solutions both in the early universe and the late time
  universe: The former greatly mitigates the coincidence problem, while the latter overcomes big rip
  disaster. The stability about the singularities is also investigated.

  Some concrete examples of  the deceleration parameter of this model
  are plotted. The result is well consistent with observations.
  Also because this model is definitely free of any extra difficulties
  from CMB decoupling, the structure formation can be studied in
  frame of it.

  We discuss the perturbation growth of this model. We find that the
  result is also consistent with observations.

 {\bf Acknowledgments.}
 We thank Z. Guo  for helpful discussions. This work was supported by
  the National Natural Science Foundation of China
    , under Grant No. 10533010, and by SRF for ROCS, SEM of China.

\end{document}